\documentclass{elsart}

\newcommand{\rd}{\mbox{{\rm d}}}
\newcommand{\bmath}[1]{\mbox{\boldmath $#1$}}
\newcommand{\fsi}{\emph{fsi}}
\usepackage{graphicx,amssymb}
\begin{document}
\begin{frontmatter}
\title{Detailed comparison of the \bmath{pp\to \pi^+pn} and
\bmath{pp\to \pi^+d} reactions at 951$\:$MeV}
\author{The GEM Collaboration:}
\author[a]{M. Abdel-Bary},
\author[d]{A. Budzanowski},
\author[i]{A. Chatterjee},
\author[g]{J. Ernst},
\author[a,c]{P. Hawranek},
\author[g]{F. Hinterberger},
\author[i]{V. Jha},
\author[a]{K. Kilian},
\author[d]{S. Kliczewski},
\author[n]{D. Kirillov},
\author[f]{D. Kolev},
\author[m]{M. Kravcikova},
\author[e]{T. Kutsarova},
\author[c]{M. Lesiak},
\author[j]{J. Lieb},
\author[a]{H. Machner}\corauth[cor]{Corresponding author}\ead{h.machner@fz-juelich.de},
\author[c]{A. Magiera},
\author[a]{R. Maier},
\author[l]{G. Martinska},
\author[a,k]{S. Nedev},
\author[o]{J. Niskanen},
\author[n]{N. Piskunov},
\author[a]{D. Prasuhn},
\author[a]{D. Proti\'c},
\author[a]{P. von Rossen},
\author[a,i]{B. J. Roy},
\author[n]{I. Sitnik},
\author[d,g]{R. Siudak},
\author[c]{M. Smiechowicz},
\author[f]{R. Tsenov},
\author[a,l]{M. Ulicny},
\author[g,l]{J. Urban},
\author[a,f]{G. Vankova},
\author[p]{C. Wilkin}
\address[a]{Institut f\"{u}r Kernphysik, Forschungszentrum J\"{u}lich, J\"{u}lich, Germany}
\address[d]{Institute of Nuclear Physics, Krakow, Poland}
\address[i]{Nuclear Physics Division, BARC, Bombay, India}
\address[g]{Institut f\"{u}r Strahlen- und Kernphysik der Universit\"{a}t Bonn, Bonn, Germany}
\address[c]{Institute of Physics, Jagellonian University, Krakow, Poland}
\address[n]{Laboratory for High Energies, JINR Dubna, Russia}
\address[m]{Technical University, Kosice, Slovakia}
\address[e]{Institute of Nuclear Physics and Nuclear Energy, Sofia, Bulgaria}
\address[j]{Physics Department, George Mason University, Fairfax, Virginia, USA}
\address[l]{P. J. Safarik University, Kosice, Slovakia}
\address[k]{University of Chemical Technology and Metalurgy, Sofia, Bulgaria}
\address[o]{Department of Physical Sciences, University of Helsinki, Finland}
\address[f]{Physics Faculty, University of Sofia, Sofia, Bulgaria}
\address[p]{Department of Physics \& Astronomy, UCL, London, U.K.}
\begin{abstract}%
The positively charged pions produced in proton-proton collisions
at a beam momentum of 1640$\:$MeV/c were measured in the forward
direction with a high resolution magnetic spectrograph. The
missing mass distribution shows the bound state (deuteron) clearly
separated from the $pn$ continuum. Despite the very good
resolution, there is no evidence for any significant production of
the $pn$ system in the spin-singlet state. However, the
$\sigma(pp\to \pi^+pn)/\sigma(pp\to \pi^+d)$ cross section ratio
is about twice as large as that predicted from $S$-wave
final-state-interaction theory and it is suggested that this is
due to $D$-state effects in the $pn$ system.
\end{abstract}
\end{frontmatter}

There is a very extensive literature on the $pp\to\pi^+d$ reaction
and many detailed analyses have been made~\cite{SAID}, but much
less is known about the production of the continuum in the
$pp\to\pi^+pn$ case. Data covering low excitation energies
generally show the strong $S$-wave final-state-interaction (\fsi)
peak corresponding to the $pn$ spin-triplet which has, as a
characteristic energy scale, the binding energy of the deuteron
($B_t=2.22\:$MeV). However, the energy resolution is generally
insufficient to identify the analogous spin-singlet \fsi\ peak,
for which the corresponding energy scale is only
$B_s=0.07\:$MeV~\cite{Machleidt01}. Indirect evidence suggests
that spin-singlet production is much weaker than that of
spin-triplet for medium energy proton beams~\cite{Hahn}, and this
is confirmed by data from the isospin-related $pp\to\pi^0pp$
reaction, though these are limited in incident momentum or energy
resolution~\cite{Jozef}. Such weak spin-singlet production accords
well with theory, because the influence of the $\Delta$-isobar is
minimal there.

A useful way of trying to extract the spin-singlet contribution is
through the comparison of the overall strengths of the cross
sections for $pn$ and deuteron final states. Using
final-state-interaction theory, F\"aldt and Wilkin derived the
extrapolation theorem which relates the normalisations of the wave
functions for $S$-wave bound and scattering states~\cite{FW}. This
has been exploited to predict the double-differential
centre-of-mass (cm) cross section for the $S$-wave spin-triplet
component in $pp\to\pi^+pn$ in terms of the cross section for
$pp\to\pi^+d$~\cite{Boudard96}:
\begin{equation}\label{equ:d_pn}
\frac{\rd^2\sigma}{\rd\Omega\,\rd{x}}
(pp\to\pi^+\left\{pn\right\}_t) = \frac{p(x)}{p(-1)}
\frac{\sqrt{x}}{2\pi(x+1)}\,\frac{\rd\sigma}{\rd\Omega}(pp\to
\pi^+d)\:.
\end{equation}
Here $x$ denotes the excitation energy $\varepsilon$ in the $np$
system in units of $B_t$, $x=\varepsilon/B_t$, and $p(x)$ and
$p(-1)$ are the pion cm momenta for the $pn$ continuum or deuteron
respectively.

In the derivation of Eq.~(\ref{equ:d_pn}) it is
assumed~\cite{Boudard96} that the pion production operator is of
short range and that $x$ is not too large, so that the $pn$
$P$-waves contribute little. Most critical though is the neglect
of channel coupling through the $pn$ tensor force, so that the
equation could only be valid provided that the $D$-state effects
are small in the production of both the bound state and continuum.

The \fsi\ peak arises from the $\sqrt{x}/(x+1)$ factor in
Eq.~(\ref{equ:d_pn}) and there should be an analogous spin-singlet
enhancement, where the deuteron binding energy $B_t$ is replaced
by the energy $B_s$ of the virtual state in the $S=0$, $T=1$
system. At low excitation energies one therefore expects that
\begin{equation}\label{equ:singlet_triplet}
\frac{\rd^2\sigma}{\rd\Omega\, \rd{x}}
(pp\to\pi^+\left\{pn\right\}_{s})= \xi\left(\frac{\varepsilon+B_t}
{\varepsilon+B_s}\right)\,\frac{\rd^2\sigma}{\rd\Omega\,\rd{x}}
(pp\to\pi^+\left\{pn\right\}_{t})\:,
\end{equation}
where we use the factor $\xi$ to quantify the ratio of
spin-singlet to spin-triplet production.

Since the best resolution in excitation energy so far achieved was
typically $\sigma=350\:$keV~\cite{Pleydon99}, any singlet peak
would have been smeared significantly in all published data.
However, by estimating the $S$-wave triplet contribution to the
$pp\to\pi^+pn$ cross section from Eq.~(\ref{equ:d_pn}) and
subtracting it from the observed data, some measure for the
singlet production could be obtained. In most experiments where
only the $\pi^+$ was detected, the limited resolution did not
guard against some leakage of the deuteron peak into the continuum
region~\cite{Gabathuler,Abaev88,Falk85}. On the other hand,
detecting the $\pi^+$ and proton in coincidence~\cite{Abaev01},
while identifying well the continuum channel, loses the relative
normalisation with the $\pi^+d$ final state, which is so important
in the implementation of Eq.~(\ref{equ:d_pn}). Therefore, in
addition to the pion spectrum, Betsch et al.~\cite{Betsch99}
measured coincidences between pion and proton, but then had to
rely on Monte Carlo simulations. For 600$\:$MeV and below, the
data seemed to confirm that the singlet contributed at most 10\%
of the cross section, though at 1$\:$GeV a higher figure was
likely~\cite{Abaev88}.

Most of the uncertainties mentioned above could be minimised by
measuring just the pion spectrum, but with high resolution. One
could then identify clearly any singlet peak and also separate
unambiguously the $pp\to\pi^+pn$ from the $pp\to\pi^+d$ reaction.
This was our primary goal when planning a new experiment. Pions
were observed near zero degrees with the 3Q2D spectrograph Big
Karl~\cite{Drochner98} at the COSY accelerator in J\"{u}lich.
Their position and track direction in the focal plane were
measured with two packs of multiwire drift chambers, each having
six layers. The chambers were followed by scintillator hodoscopes
that determined the time of flight over a distance of $3.5\:$m. In
order to optimise the momentum resolution, a liquid hydrogen
target of only $2\:$mm thickness was used with windows made of
$1\:\mu$m Mylar~\cite{Jaeckle94}. The beam was electron cooled at
injection energy and, after acceleration, stochastically
extracted. This resulted in an energy resolution of
$\sigma=97\:$keV for the deuteron peak. This was much better than
that found in a test run without beam cooling and, in particular,
the background was considerably reduced.

\begin{figure}
\begin{center}
\includegraphics[width=10 cm]{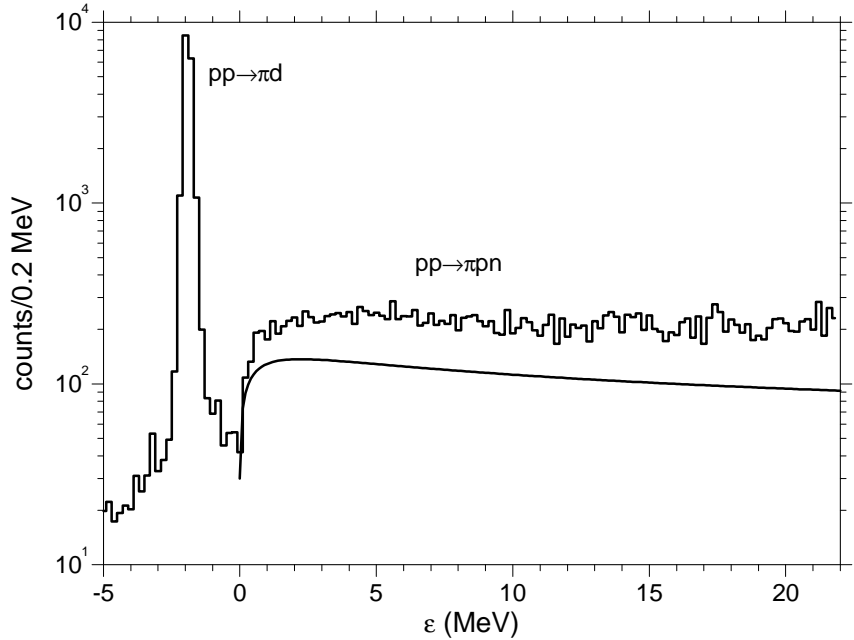}
\end{center}
\caption{The results from the present experiment (histogram)
compared with the prediction (curve) of the $S$-wave \fsi\ theory
of Eq.~(\ref{equ:d_pn})~\protect\cite{Boudard96}.}
\label{FW_model}
\end{figure}

The results of our experiment are shown in Fig.~\ref{FW_model} as
function of the excitation energy in the $pn$ system. Though
corrections for acceptance \emph{etc.}\ have been included, these
in fact vary slowly with $\varepsilon$ for energies below
$20\:$MeV. Noting the logarithmic scale in the figure, it is clear
that there is an excellent distinction between the $pp\to\pi^+pn$
from the $pp\to\pi^+d$ reactions. Since the luminosity and
detection efficiencies largely cancel out between them, this means
that we have a very good determination of the relative cross
sections for $\pi^+d$ and $\pi^+pn$ final states.

Also shown in Fig.~\ref{FW_model} is the prediction of the
continuum production from the $S$-wave \fsi\ theory of
Eq.~(\ref{equ:d_pn}), where we have assumed a constant background
of 30 counts per bin. Though the shape is largely right, it is too
low in magnitude by a factor of $2.2\pm 0.1$ over the whole of the
spectrum. It is interesting to note that, if our data are
artificially degraded such that the resolution is the same as that
achieved in the Leningrad experiment at the neighbouring energy of
1$\:$GeV ($\sigma\approx 3\:$MeV)~\cite{Abaev88}, the two sets of
results overlap very well. However, the poor resolution allowed
the authors of ref.~\cite{Boudard96} to ascribe the factor-of-two
discrepancy to the production of spin-singlet final states. We
can, however, check this hypothesis independently by studying the
shape of the missing-mass spectrum.

\begin{figure}\centering
\includegraphics[width=10 cm]{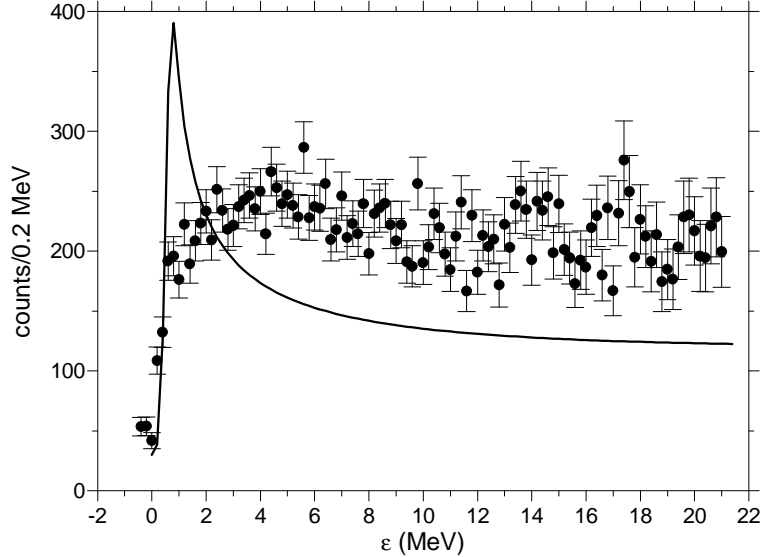}
\caption{Comparison of the measured $pn$ excitation energy
spectrum on a linear scale with the prediction of
Eqs.~(\ref{equ:d_pn},\ref{equ:singlet_triplet}) for the shape of
the singlet cross section. The error bars contain a tiny
contribution from the uncertainty in the acceptance correction. }
\label{singlet}
\end{figure}

As is evident from Eq.~(\ref{equ:singlet_triplet}), the cross
section for producing a $pn$ singlet state must show a sharp spike
just above threshold and, due to our good resolution, this prominent
feature should remain even after convolution with this resolution.
In Fig.~\ref{singlet} are shown the predictions of
Eqs.~(\ref{equ:d_pn},\ref{equ:singlet_triplet}) modified by the
inclusion by an extra factor of $(1+\varepsilon/E_s)$ to try to take
into account deviations from the extrapolation theorem~\cite{FW}.
The value of $E_s=24\:$MeV is derived from the scattering length and
effective range~\cite{GW} though, by the point that this becomes
significant, the $S$-wave \emph{ansatz} is dubious. There is no hint
of any sharp needle in the data of Fig.~\ref{singlet} and, in fact,
the shape of the cross section is completely compatible with pure
spin-triplet production. Fits of Eq.~(\ref{equ:singlet_triplet})
with free amounts of singlet and triplet show that $\xi < 10^{-4}$
at the one standard deviation level, and this corresponds to a
practically  vanishing fraction if the singlet part. As a
consequence, we must seek elsewhere for the factor-of-two
discrepancy between our data and the results of
Eq.~(\ref{equ:d_pn}).

As has been stressed previously, the extrapolation theorem linking
the bound and scattering wave functions is only valid if one can
neglect completely $D$-state effects~\cite{FW}. Though the
$D$-state wave functions are suppressed at short distances by the
centrifugal barrier, the $S$-wave is also reduced in this region
by the repulsive core. Thus the $D$-state might be significant for
pion production despite the relatively small probability in the
deuteron, especially if $S$-$D$ interference terms are important.

We consider a microscopic calculation of the actual three-body
$\pi^+pn$ final state reaction to be beyond the scope of the
present work. Nevertheless, to investigate the effects of the
$D$-wave, at least semi-quantitatively, we have made estimations
of the $pp\rightarrow \pi^+d$ differential cross section following
the formalism described in ref.~\cite{Jouni}. Using a standard
deuteron wave function~\cite{Reid} with a normal $D$ state, this
reproduces well the experimental data~\cite{SAID}. The
calculations have, however, been repeated with a reversed sign for
the $D$-state amplitude and also with no $D$-state at all. Now for
kinematic reasons the $pn$ $D$-state scattering wave function must
vanish like $\varepsilon^1$ as $\varepsilon\to 0$ so that its sign
should change when going from the bound state (deuteron) to the
continuum $pn$ pair~\cite{FW}. One can therefore get an idea of
the effect of the $D$-state in the continuum by using a deuteron
wave function with the opposite sign for the $D$ wave.

\begin{figure}
\begin{center}
\includegraphics[width=10 cm ,angle=0]{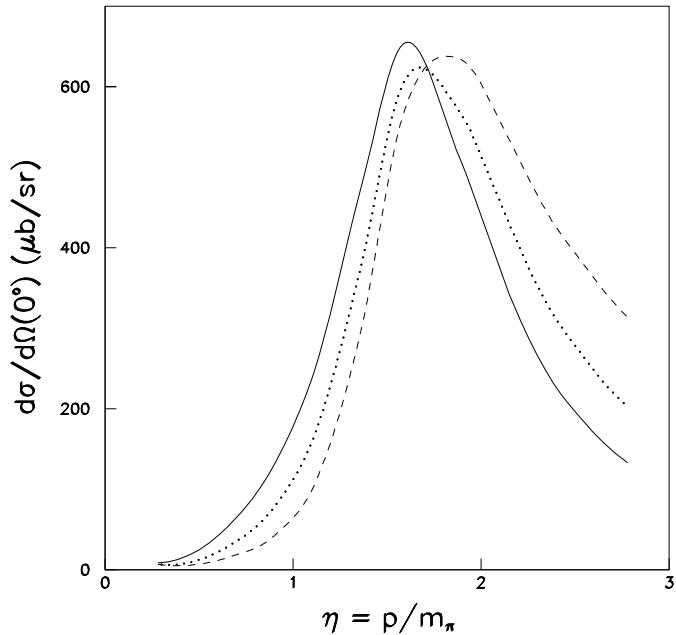}
\caption{$D$-state effects in the predicted excitation
function~\protect\cite{Jouni} for the zero degree $pp\to \pi^+d$
differential cross section. The solid curve shows the results with
the standard value~\protect\cite{Reid}, the broken curve with the
reversed sign, and the dots with no $D$-state at all.}
\label{d_State}
\end{center}
\end{figure}

The predictions for the forward cross section are shown in
Fig.~\ref{d_State} as a function of the dimensionless pion cm
momentum $\eta = p/m_{\pi^+}$, the present experiment
corresponding to $\eta=2.6$.  The zero $D$-state calculation is
approximately the average of the other two, showing that the
effects are mainly due to $S$-$D$ interference. At low energies
the inclusion of the $D$ state increases the cross section, while
the converse is true at high energies. The exact position of the
cross-over point, here predicted to be at $\eta\approx 1.6$
($T_p\approx 600\:$MeV), could be model dependent but we would
certainly expect there to be a different influence of the
$D$-state on either side of the $\Delta$ peak. Given the
theoretical uncertainties, the fact that the factor of 2.2
difference between the calculations with the changed sign of the
$D$-state at $\eta=2.6$ coincides exactly with the discrepancy
between the data and the $S$-wave theory shown in
Fig.~\ref{FW_model} may be fortuitous. Close to or just below the
resonance one would expect smaller deviations from the
extrapolation theorem associated with the $D$-state, and this
certainly seems to be the case
experimentally~\cite{Boudard96,Pleydon99,Gabathuler,Falk85}. To
quantify the deviations would require further high resolution runs
which could identify clearly the singlet production from the shape
of the spectrum.

In summary, we have measured the missing mass spectrum from the
$pp\to \pi^+ X$ reaction in the forward direction. Despite the
rather high beam momentum of 1640$\:$MeV/c, the excellent
resolution allowed the complete separation of the deuteron from
$pn$ continuum and also showed that the production of spin-singlet
states was negligible at this momentum. Deviations from the
results of $S$-wave \fsi\ theory could be ascribed
semi-quantitatively to the effects of the tensor force in the $pn$
system and an extension of this to encompass the coupled $S$-$D$
system would be of great help. It is also to be hoped that a full
microscopic calculation of the three-body $\pi^+pn$ final state
production will be undertaken to complement the two-body results
quoted here~\cite{Jouni}. This might then confirm our hypothesis
of the great influence of the deuteron $D$-state in pion
production above the $\Delta$ resonance.
\\[1ex]

The quality of the beam necessary for the success of this work is
due mainly to the efforts of the COSY operator crew. Support by BMBF
Germany (06 MS 568 I TP4), Internationales B\"uro des BMBF (X081.24
and 211.6), SGA Slovakia (1/1020/04), Academy of Finland (54038),
and the Forschungszentrum J\"ulich is gratefully acknowledged.



\begin{thebibliography}{99}
%
\bibitem{SAID} R.A.~Arndt et al., Phys.\ Rev.\ C 48 (1993) 1926,
\mbox{http://gwdac.phys.gwu.edu/}
%
\bibitem{Machleidt01} R.~Machleidt, Phys.\ Rev.\ C 63 (2001) 024001.
%
\bibitem{Hahn} H.~Hahn et al., Phys.\ Rev.\ C 53 (1996) 1074.
%
\bibitem{Jozef} See for example: R.~Bilger et al., Nucl.\ Phys.\ A 663
(2001) 633.
%
\bibitem{FW} G.~F\"{a}ldt and C.~Wilkin, Phys.\ Scripta 56 (1997)
566.
%
\bibitem{Boudard96}
A.~Boudard, G.~F\"{a}ldt and C.~Wilkin: Phys.\ Lett.\ B 389, 440
(1996).
%
\bibitem{Pleydon99}
R.G.~Pleydon et~al., Phys.\ Rev.\ C 59 (1999) 3208.
%
\bibitem{Gabathuler} K.~Gabathuler et al., Nucl.\ Phys.\ B 40
(1972) 32.
%
\bibitem{Abaev88} V.V.~Abaev et al., Leningrad report LNPI-80-569
(unpublished); J.\ Phys.\ G 14 (1988) 903.
%
\bibitem{Falk85} W.R.~Falk et al., Phy.\ Rev.\ C 32 (1985) 1972.
%
\bibitem{Abaev01}
V.V.~Abaev et~al., Phys.\ Lett.\ B 521 (2001) 158; Y.N.~Uzikov and
C.~Wilkin, Phys.\ Lett.\ B 511 (2001) 191.
%
\bibitem{Betsch99}
A.~Betsch et~al.: Phys.\ Lett.\ B 446 (1999) 179.
%
\bibitem{Drochner98}
M.~Drochner et~al., Nucl.\ Phys.\ A 643 (1998) 55.
%
\bibitem{Jaeckle94}
V.~Jaeckle et~al., Nucl.\ Instrum. Methods A 349 (1994) 15.
%
\bibitem{GW} M.L.~Goldberger and K.M.~Watson, \emph{Collision
Theory}, Wiley (New York, 1964), sect.~9.3.
%
\bibitem{Jouni} J.~Niskanen, Nucl.\ Phys.\ A 298 (1978) 417;
Phys.\ Rev.\ C 49 (1994) 1285.
%
\bibitem{Reid} R.~Reid, Ann.\ Phys.\ (N.Y.) 50 (1968) 411.
%
\end{thebibliography}
\end{document}